\documentclass[twocolumn,usenatbib]{aa}
\usepackage{epsfig,natbib}
\def\s5{S5~0716+714}

\newcommand{\lsim}{{\lower.5ex\hbox{$\; \buildrel < \over \sim \;$}}}
\newcommand{\gsim}{{\lower.5ex\hbox{$\; \buildrel > \over \sim \;$}}}

\def\ee{\end{equation}}
\def\be{\begin{equation}}
\begin{document}

\title{AGILE and Swift simultaneous observations of the blazar S50716+714 during the bright flare of October 2007}
\author{P.~Giommi\inst{1}, S.~Colafrancesco\inst{1}, S. Cutini\inst{1}, P. Marchegiani\inst{2}, M.Perri\inst{1}, C. Pittori\inst{1}, F. Verrecchia\inst{1},  A. Bulgarelli\inst{3}, A. Chen\inst{4}, F. D'Ammando\inst{5}, I. Donnarumma\inst{5}, A. Giuliani\inst{4}, F. Longo\inst{6}, L. Pacciani\inst{5},\\
G. Pucella\inst{5}, S. Vercellone\inst{4}, V. Vittorini\inst{5}, M. Tavani\inst{5}
}
\institute{
            ASI Science Data Center, ASDC c/o ESRIN,
            via G. Galilei snc I-00044 Frascati, Italy.
            \and
            Dipartimento di Fisica,   Universit\`a  Sapienza, P.le Aldo Moro, 2, I-00185 Roma, Italy
            \and
            INAF-IASF Bologna, via  Gobetti, 101, I-40129 Bologna, Italy
            \and
           INAF, IASF-Milano
            via E. Bassini 15, I-20133, Milano, Italy
            \and
            INAF, IASF-ROMA,
            via Fosso del Cavaliere 100, I-00133, Roma, Italy
            \and
            INFN Trieste, via Padriciano 99, I-34012 Trieste, Italy
}

\date{Received ; Accepted }

\authorrunning{Giommi et al.}
\titlerunning{AGILE and Swift observations of the blazar S50716+714 in October 2007}
\vspace{-0.2cm}
\abstract{We present the results of a series of optical, UV, X-ray and
$\gamma$-ray observations of the BL Lac object S50716+714 carried out by
the Swift and AGILE satellites in late 2007 when this blazar was flaring
 close to its historical maximum at optical frequencies.
 We have found that the optical through soft X-ray emission, likely due to
 Synchrotron radiation, was highly variable and displayed a different
 behavior in the optical UV and soft X-ray bands. The 4-10 keV flux, most
 probably dominated by the inverse Compton component, remained instead
 constant. The counting statistics in the relatively short AGILE GRID
 observation was low and consistent with a constant $\gamma$-ray flux
 at a level similar to the maximum observed by EGRET.
 An estimate of the $\gamma$-ray spectral slope gives a value of the
 photon index that is close to 2  suggesting that the peak of the inverse
 Compton component in the Spectral Energy Distribution (SED) is within the
 AGILE energy band.  The different variability behavior observed in
 different parts of the SED exclude interpretations predicting highly
 correlated  flux variability like changes of the beaming factor or of the
 magnetic field in simple SSC scenarios.
The observed SED changes may instead be interpreted as due to the sum
of two SSC components, one of which is constant while the other is
variable and with a systematically higher synchrotron peak energy.}
\maketitle


\vspace{-0.5cm}
\section{Introduction}

\vspace{-0.5cm}
Blazars are a rare class of radio loud AGN known to display extreme properties like irregular, often large
and rapid variability, apparent super-luminal motion, flat radio spectrum, large and variable
polarization.
Blazars are thought to be objects emitting
radiation over the entire electromagnetic spectrum from a relativistic jet that is viewed closely along
the line of sight thus causing strong relativistic amplification (e.g., \cite{Urry95}).
Blazars come in two flavors:  Flat Spectrum Radio Quasars \citep[FSRQs, or blazars of the Q type in the nomenclature of][]{gio_wmap_08}, and
BL Lac objects,  \citep[or  blazars of the B type  in][] {gio_wmap_08} depending on whether their optical spectrum does or does not show the strong and broad emission lines that are typical of QSOs.
The broad-band continuum emission of both types of blazars is normally associated to synchrotron and inverse Compton emission, which is expected to produce the two-bump shaped Spectral Energy Distribution (SED) that is always observed in these sources. The synchrotron component usually peaks [in a $Log(\nu f(\nu))-Log(\nu)$ representation] in the infra-red or in the optical band but is some cases, almost exclusively in Blazars of the BL Lac type, the synchrotron emission can
peak at much higher energies, up to the hard X-ray band.

\s5 is an optically bright and well studied BL Lac known to be highly variable
at all frequencies \citep[e.g.][]{wagner96, Gio99a} and on timescales ranging
from years to a few thousand seconds
\citep[e.g.][]{raiteri03,nesci05,foschini06}. Within the classification of
\cite{P95} \s5 is an intermediate object (IBL) between Low Energy Peaked (LBL)
and High Energy Peaked (HBL) blazars since its synchrotron component reaches
the soft X-ray and the Inverse Compton component has been detected in the
hard-X-rays \citep{Gio99a, Tagliaf03}. Despite the many attempts to determine
its redshift no positive detection of any feature in its optical spectrum has
ever been reported. This, combined with the lack of any host galaxy sets a
lower limit of 0.52 to its redshift \citep{sbarufatti05}. \s5 was detected
several times by  the EGRET $\gamma$-ray detector of the Compton Observatory
\citep{Hartman99}.

In this {\it Letter} we present a series of AGILE and Swift ToO observations of
\s5 carried out  between 23 October and 13 November 2007 following an AGILE
detection of  this source with a large $\gamma$-ray flux in September 2007
(Giuliani et al. 2007, ATEL 1221) and an alert issued in late October 2007 by
an optical monitoring campaign reporting that this source was undergoing a
large optical flare  \citep[see][for details]{Villata_0716_08}.

\vspace{-0.4cm}
\section{AGILE observations and data analysis}

The AGILE observatory \citep{tavani08} is a new technology satellite of the
Italian Space Agency (ASI) dedicated to $\gamma$-ray and X-ray astrophysics.
AGILE was successfully launched on 23 April 2007 into a low Earth, nearly
equatorial, orbit which ensures a low particle background and short passes
through the South Atlantic Anomaly. AGILE is capable of observing cosmic
sources simultaneously in the 18 Ð- 60 keV and 30 MeV -Ð 50 GeV energy bands
with the SuperAGILE (SA), \citep{feroci07} and the Gamma Ray Imaging Detector
(GRID)  respectively. The GRID is a combination of a Silicon Tracker
\citep{prest03, barbiellini01}, a Mini-Calorimeter \citep{labanti06} and an
Anti-coincidence System \citep{perotti06}

On 23 October 2007, \s5 was reported by optical observers to be flaring at a
flux close to its historical maximum \citep{Villata_0716_08}. Despite AGILE was
still undergoing its in-flight calibration program, on 24 October the satellite
began a ToO maneuver  to bring \s5 in its very large field of view ($\approx $
60 deg radius).
The ToO lasted until 1 November 2007 11:50 UTC, for a total pointing duration
of  $\sim $ 8 days.

All GRID data are normally processed using the scientific data reduction
software developed by the AGILE instrument teams and integrated into an
automatic pipeline system developed at the ASI Science Data Center (ASDC). The
8 days of  \s5 October ToO data have been processed with the most recent
software and in-flight calibrations available at the time of writing
($BUILD\_GRID$ 14). The AGILE software was run creating counts, exposure, and
galactic background $\gamma$-ray  maps with a bin-size of $0.3 \times 0.3$ deg.
To reduce the particle background contamination only events tagged as confirmed
$\gamma$-ray events were selected (F4 filtercode=5),
All the $\gamma$-ray events whose reconstructed directions form angles with the
satellite-Earth vector smaller than 80 deg (albrad=80) were also rejected, thus
reducing the $\gamma$-ray Earth Albedo contamination.  We created maps in the
full AGILE energy band  E $>$ 100 \rm{MeV},  and in the two separate energy
bands: 100-400 \rm{MeV} and 400-1000 \rm{MeV}.
The AGILE Maximum Likelihood procedure gives the average flux values reported
in Table \ref{tab.GRIDresults}
\noindent
\begin{table}[h*]
\caption{Results of AGILE-GRID analysis}
\begin{tabular}{cccc}
\hline \multicolumn{1}{c}{Energy band} & \multicolumn{1}{c}{Count rate} &
\multicolumn{1}{c}{Net counts}& \multicolumn{1}{c}{  $\sqrt{TS}$}\\
 \multicolumn{1}{c}{} &
 \multicolumn{1}{c}{($\rm{ph~cm^{-2} s^{-1}}$) } &
  \multicolumn{1}{c}{} &
 \multicolumn{1}{c}{}\\
\hline
$>$ 100~MeV& (4.9$\pm$1.1)$\times 10^{-7}$&48$\pm$11&6\\
100-400~MeV& (2.3$\pm$0.9)$\times 10^{-7}$ &28$\pm$10&3\\
400-1000~MeV& (6$\pm$2)$\times 10^{-8}$ &11$\pm$4&4\\
\hline
\end{tabular}
\label{tab.GRIDresults}
\end{table}

A preliminary spectral fitting based on a least square fit to the flux
estimated in different bands gives a photon spectral index of $\sim 1.9 \pm
0.5$. The low counting statistics  in the rather short (8 days) AGILE-GRID
observation does not allow us to detect small amplitude variability. The
overall flux is consistent with a constant value or with variability of
magnitude less than a factor of $\approx$3. A description of the  \s5 GRID
observations and further details of the data analysis are reported in
\citet{chen08}. \s5 is too faint  to be detected by the Super-AGILE X-ray
detector.

\vspace{-0.2cm}
\section{Swift observations}

Swift performed 10 short ToO observations of \s5 between 23 October and 13 November 2007
as detailed in Table \ref{tab.logswift}.
The observations were carried out using all three Swift on-board experiments:
the X-ray Telescope \citep[XRT,][]{Burrows05}, the UV and Optical Telescope
\citep[UVOT,][]{Roming05} and the Burst Alert Telescope \citep[BAT,][]{Barthelmy05}. However, the hard-X-ray flux of \s5 is below
the sensitivity of the BAT instrument for short exposure and therefore the data from this instrument will not be used here.

\vspace{-0.2cm}
\begin{table}[h*]
\caption{Log of Swift observations in October-November 2007}
\begin{tabular}{ccc}
\hline
\multicolumn{1}{c}{Observation date} &
\multicolumn{1}{c}{XRT exposure} &
\multicolumn{1}{c}{Count rate}\\
 \multicolumn{1}{c}{ } &
 \multicolumn{1}{c}{seconds} &
 \multicolumn{1}{c}{ct/s 0.3-10 keV}\\
\hline
23-Oct-2007& 2443&0.47$\pm$0.015\\
24-Oct-2007& 1991&0.29$\pm$0.013 \\
25-Oct-2007& 2698&0.37$\pm$0.013\\
26-Oct-2007&  2199& 0.34$\pm$0.013\\
27-Oct-2007&  1961&0.72$\pm$0.030\\
28-Oct-2007& 1758 &0.33$\pm$0.015\\
03-Nov-2007&  2801&0.90$\pm$0.030\\
06-Nov-2007&  1923&0.28$\pm$0.013\\
09-Nov-2007&  2003&0.39$\pm$0.015\\
13-Nov-2007&  2382&0.27$\pm$0.012\\
\hline
\end{tabular}
\label{tab.logswift}
\end{table}

\vspace{-0.8cm}
\subsection{UVOT data analysis}

During all observations UVOT produced a series of images in each of the lenticular filters (V, B, U, UVW1, UVM2, and UVW2).
Photometry of \s5 was performed using the standard UVOT software distributed within the HEAsoft 6.3.2 package and the
calibration included in the latest release of the ``Calibration Database''.
Counts were extracted from aperture of 5\arcsec\, radius for all filters and converted to fluxes using the standard zero points.
The fluxes were then de-reddened using a value for $E(B-V)$ of 0.031 mag \citep{schlegel1998} with $A_{\lambda}/E(B-V)$ ratios calculated for UVOT filters using the mean interstellar extinction curve from \citet{Fitzpatrick1999}. No variability was detected within
single exposures in any filter.
Part of the UVOT data presented here also appear in \citet{Villata_0716_08} who reported results fully consistent with ours.

\vspace{-0.2cm}
\subsection{XRT data analysis}

All XRT observations were carried using the instrument in Photon Counting readout  mode, which provides maximum
sensitivity but is affected by photon pile up for count-rates larger than $\approx 0.5 $ cts/s.
The data were reduced using the {\it XRTDAS} software developed at the ASI/ASDC and distributed within the NASA/HEASARC HEAsoft 6.3.2 package. Photons were selected with grades in the range 0-12 and default screening parameters were used to produce level 2 cleaned event files. Spectral data were extracted in a circular region with a radius of 20 pixels.
To avoid pile-up problems, when the source count rate was larger than 0.5 cts/s, the spectral data were extracted in an annular region with inner and outer radii of 3 and 20 pixels respectively 
The background was estimated in a nearby source-free circular region of 50
pixels radius. In order to use $\chi^2$ statistics, spectra were rebinned to
include at least 20 photons in  each energy channel. For our spectral fitting
we used the XSPEC 11.3 spectral analysis package. Given the very short
exposures (typically $\approx$ 2,000 seconds, see Tab \ref{tab.logswift}) the
limited photon statistics allowed us to fit the data to a simple power law
spectral model with low energy absorption ($N_H$) fixed to to the Galactic
value of $3.8\times 10^{20}$ cm$^{-2}$ as estimated by \cite{Dick90}.
\begin{figure}[!ht]
\begin{center}
\includegraphics[height=7.4cm,angle=-90]{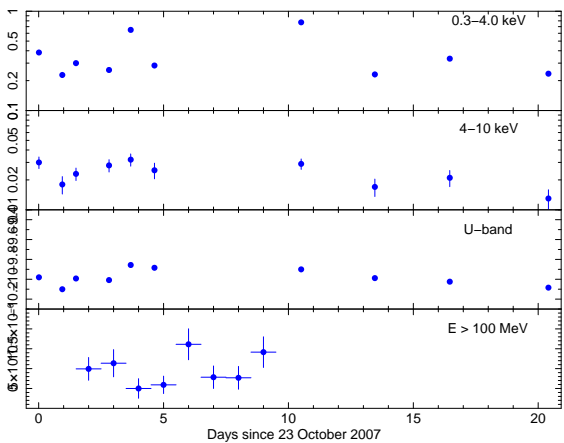}
\caption{ The light curves of S50716+714 in the soft XRT band  (0.3-4 keV, top panel),
 in the hard XRT band (4-10 keV, second panel from top), in the UVOT U filter (third panel) and
 in the AGILE-GRID detector (bottom panel).
 The vertical axis of the Swift light curves is logarithmic with one decade range for all cases, showing that variations
 are largest in the soft X-ray band.}
 \end{center}
 \label{fig.LC}
\end{figure}

The results are reported in Tab. \ref{tab.spectral_results} where column 1
gives the observation date, column 2 gives the flux in the 2-10 keV band,
column 3 gives the power law photon index [$f (ph~cm^{-2}s^{-1}) \propto
~$E$^{-\Gamma}$]  and 1$\sigma$ error, and column 4 gives the reduced $\chi^2$
and the number of degrees of freedom. In all cases the average photon index is
steeper than 2, indicating that the 0.3-10 keV X-ray spectrum  is dominated by
synchrotron radiation. However, from the X-ray light curves of Fig. 1 clearly
show that the soft (0.3-4 keV) X-ray flux varies much more than the hard (4-10
keV) X-ray flux, suggesting that the high energy part of the spectrum is mostly
due the rise of the inverse Compton component as is also apparent in the
broad-band SED of Fig. 2.

\vspace{-.4cm}
\begin{table}[h*]
\caption{Results of XSPEC fits. Power law model with $N_{\rm H}$ fixed to the
Galactic value of $3.8\times 10^{20}$ cm$^{-2}$. }
\begin{tabular}{cccc}
\hline
\multicolumn{1}{c}{Observation} &
\multicolumn{1}{c}{2-10 keV flux} &
 \multicolumn{1}{c}{Spectral slope} &
 \multicolumn{1}{c}{Reduced} \\
 \multicolumn{1}{c}{date}&
 \multicolumn{1}{c}{erg cm$^{-2}$ s$^{-1}$} &
 \multicolumn{1}{c}{$\Gamma$}&
 \multicolumn{1}{c}{$ \chi^2$/d.o.f.}\\
\hline
23-Oct-2007 & 5.9$\times 10^{-12}$& $2.3 \pm 0.05$ &0.74/40\\
24-Oct-2007 & 4.5$\times 10^{-12}$& $2.2 \pm 0.08$ &0.55/18\\
25-Oct-2007 & 4.8$\times 10^{-12}$ & $2.2 \pm 0.05$ &0.81/35\\
26-Oct-2007 & 5.1$\times 10^{-12}$ & $2.1 \pm 0.07$ &1.37/26\\
27-Oct-2007 & 7.6$\times 10^{-12}$ & $2.4 \pm 0.10$ &0.68/32\\
28-Oct-2007 & 5.6$\times 10^{-12}$ & $2.2 \pm 0.07$ &0.87/19\\
03-Nov-2007 & 8.0$\times 10^{-12}$ & $2.6 \pm 0.07$ &0.98/54\\
06-Nov-2007 & 3.9$\times 10^{-12}$ & $2.4 \pm 0.08$ &1.28/17\\
09-Nov-2007 & 5.6$\times 10^{-12}$ & $2.4 \pm 0.07$ &1.08/24\\
13-Nov-2007 & 3.7$\times 10^{-12}$ & $2.4 \pm 0.07$ &0.92/21\\
\hline
\end{tabular}
\label{tab.spectral_results}
\end{table}

\vspace{-0.8cm}
\section{Discussion}

We have presented a series of simultaneous ToO observations of the blazar \s5
carried out by the Swift and AGILE satellites following the detection of this
source in a state of large flaring activity \citep{Villata_0716_08}.
\begin{figure}[!ht]
\begin{center}
\includegraphics[height=7.1cm,angle=-90]{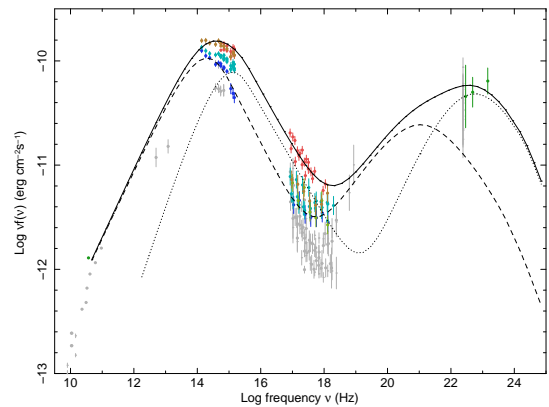}
\caption{ The SED of S50716+714 built combining our  data with simultaneous
optical and radio data from \citet{Villata_0716_08} (color) and with archival
multi-frequency data (light grey). The dashed and dotted lines represent the
SSC components described in the text.}
 \end{center}
 \label{fig.SED}
\end{figure}

The light-curves shown in Fig. 1 show strong variability (up to a factor
$\approx 4$) in the soft-X-ray flux, more moderate variations (less than a
factor 2) at optical/UV frequencies, and an approximately constant hard X-ray
(5-10 keV) flux (see also Fig. 2). A similar behavior was seen during a {\it
Beppo}SAX observation of \s5 carried out in October 2000
\citep{Tagliaf03,foschini06,ferrero06}. The low counting statistics accumulated
during the relatively short ($\sim8$ days) AGILE observation does not allow to
detect small amplitude variability. The count-rate is consistent with a
constant flux or with variations  lower than a factor of $\approx$ 3.

The Spectral Energy Distribution of \s5 built with all the Swift 
and AGILE data is shown in Fig. 2 where we see that our flux measurements
(colored filled circles) are much larger than those from archival data (light
grey points). Large intensity and spectral changes close to the peak and at the
end of the synchrotron component, are clearly present while the rise of the
inverse Compton component above approximately 4 KeV remained constant over the
entire observing period. The AGILE GRID data, integrated between 23 October and
1 November, is shown as three points between $\approx $100 MeV and 1 GeV which
have been derived from the fluxes reported in table \ref{tab.GRIDresults}. The
intensity is close to the maximum observed by EGRET (grey points). This flaring
behavior is quite different from that seen in 3C454.3 in April/May 2005 where
the inverse Compton component varied by a large factor in a few days
\citep{giommi06}. This demonstrates that flaring events in blazars may be
widely different and each should ideally be followed with detailed observations
covering the dynamical times scales of all the emission components. The
observed spectral behavior of \s5 cannot be the result of simple changes in the
magnetic field or of the beaming factor in a one-zone homogeneous SSC model. In
this case the emission in different bands is expected to vary in a highly
correlated way, contrary to what is observed in the optical and X-ray data  of
Fig. 1 and in the well sampled simultaneous light curve of
\citet{Villata_0716_08} where the R-band variations are considerably smaller
then in the soft X-ray band. This behavior would also challenge external
Compton model \citep{Sik94,Tagliaf03} in which variable high energy emission
should result from a single electron population accelerated by disturbances in
the jet.

The different flux variations observed at optical and X-ray frequencies could
instead be explained by the presence of two SSC components, one of which is
constant over the entire observing period, while the second one is highly
variable and possibly due to a secondary blob of relativistic plasma including
fresh and more energetic particles (see Fig. 2). However, even this scenario
cannot reproduce exactly the light curves and spectral changes in all energy
bands as this would require a very good knowledge of all the acceleration
events and a proper treatment of the energy-dependent loss mechanisms. This can
be achieved only through a continuous monitoring of the source while our
optical, UV and X-ray observations are just snapshots of $\approx$ 2,000
seconds taken approximately at one day intervals. Fig. 2 shows the two SSC
components as dashed and dotted lines, while the continuous line is their sum.
For their evaluation we have assumed a SSC model with broken power law spectrum
for both particle components:
\begin{eqnarray}
n_e(\gamma)&=&k_0\gamma^{-p_1} \;\;\;\;\gamma\leq \gamma_b\\
n_e(\gamma)&=&k_0\gamma_b^{-p_1}(\gamma/\gamma_b)^{-p_2}
\;\;\;\; \gamma>\gamma_b,
\end{eqnarray}
with constant density within a spherical region of radius $r$ and with $\gamma$ in the range $10^2 - 10^6$.
The parameters of the "stationary" component, which
reproduces the "minimum" fluxes at UV and X-ray frequencies,
are $p_1=1.7$ and $p_2=4.2$ at low and
high energies respectively, and the break Lorentz factor is
$\gamma_b=10^{3.25}$; for the second component  similar parameters are
assumed ($p_1=1.1$ and $p_2=4.05$), with a larger break Lorentz
factor of $\gamma_b=10^{3.6}$, which produces a stronger emission at high energies.
The magnetic field and the beaming factor for the two components
are assumed to be the same ($B=1.1$ G and $\delta=20$),
while the first component has an higher normalization and
a larger radius ($k_0=631.$ cm$^{-3}$ and $r=0.010$ pc) w.r.t.
the second component ($k_0=8.91$ cm$^{-3}$ and $r=0.005$ pc). $B$ and $r$ values are co-moving with the emitting regions.

Finally we note that during a recent large flare \citep{giommiATEL08} \s5 was
detected at TeV energies \citep{teshimaTeV08} despite its likely high redshift.
If the limit of z $>$ 0.52 \citep{sbarufatti05} is confirmed this detection
would imply that the Universe is much more transparent to TeV photons than
previously claimed.

\begin{acknowledgements}
We acknowledge funding from ASI grant I/024/05/1. This work is partly based on
multi-frequency archival data taken from the ASDC, a facility of  ASI.
\end{acknowledgements}


\end{document}